\documentclass[twocolumn,showpacs,amsmath,amssymb,aps,prl,floatfix]{revtex4}
\usepackage{graphicx}
\usepackage{dcolumn}
\usepackage{bm}

\begin{document}

\title{On the sample size dependence of the critical current density in MgB$_2$ superconductors}

\author{M. J. Qin}
\author{S. Keshavarzi}
\author{S. Soltanian}
\author{X. L. Wang}
\author{H. K. Liu}
\author{S. X. Dou}
\affiliation{Institute for Superconducting and Electronic Materials,
 University of Wollongong, Wollongong NSW 2522, Australia}

\date{\today}

\begin{abstract}
Sample size dependent critical current density has been observed in magnesium diboride superconductors. At high fields, larger samples provide higher critical current densities, while at low fields, larger samples give rise to lower critical current densities. The explanation for this surprising result is proposed in this study based on the electric field generated in the superconductors. The dependence of the current density on the sample size has been derived as a power law $j\propto R^{1/n}$ ($n$ is the $n$ factor characterizing $E-j$ curve $E=E_c(j/j_c)^n$). This dependence provides one with a new method to derive the $n$ factor and can also be used to determine the dependence of the activation energy on the current density.
\end{abstract}
\pacs{74.25.Qt, 74.25.Ha, 74.25.Op, 74.25.Sv}
\maketitle

It has been reported that the critical current density derived from magnetic measurement in the magnesium diboride superconductors depends on the sample size \cite{horvat,haojin}. Usually a larger sample results in a higher critical current density at high fields. If this is a true intrinsic property of this new superconductor, it would be advantageous to use this superconductors in large scale applications. As far as we know, such a phenomenon has not been observed in either high temperature or low temperature superconductors. Explanations have been proposed to account for this observation \cite{horvat,haojin}. Jin et al. \cite{haojin} measured the relaxation of cylindrical magnesium diboride superconductors of different lengths and found that the activation energy depends linearly on the length of the sample up to 1 $mm$ and saturates after that. The authors suggested that the vortices in the magnesium diboride superconductors are quite rigid at small sample lengths and break into segments as the sample length reaches the collective pinning length $L_c\approx(\epsilon_0^2\xi^2/\gamma)^{1/3}$, with $\epsilon_0$ the basic energy scale, $\xi$ the coherence length, and $\gamma$ a parameter of disorder strength. However, their proposal cannot explain the higher critical current density for smaller samples at low fields. Horvat et at. \cite{horvat} proposed that different coupling between the superconducting grains at different length scales is responsible for the sample size dependent critical current density. However, this explanation is quite qualitative and not conclusive. 

It is very important to clarify this problem. On the one hand, we need to understand the underlying mechanism governing this dependence in order to see whether we can further improve the critical current density by increasing the sample size and to understand why this phenomenon has not been reported in high temperature or low temperature superconductors. On the other hand, we need a standard to compare the current density of magnesium diboride superconductors fabricated by different techniques. In this paper, we propose an explanation for this observation based on the electric field generated in the superconductors during a hysteresis loop measurement. 

The samples used in this study are all in the shape of rectangular rods cut from a MgB$_2$ pellet. The sample preparation can be found elsewhere \cite{wang}. In order to eliminating any geometric effects on the critical current density, the pellet was cut into a series of samples with constant size ratio $a:b:c$. Seven samples are used in this study with dimensions of $a\times b\times c$ $(mm^3)$: $1.07\times 3.27\times 7.15$, $0.7\times 2.12\times 4.65$, $0.57\times 1.68\times 3.64$, $0.46\times 1.34\times 2.92$, $0.36\times 1.08\times 2.29$, $0.29\times 0.85\times 1.87$, $0.24\times 0.68\times 1.42$ $mm^3$. The critical current density is derived from magnetic hysteresis loop measurements by means of a Quantum Design PPMS (Physical Property Measurement system) magnetometer with a sweep rate of 50 Oe/s. The measurements are performed with the applied field parallel to the longest direction of the sample ($c$ axis). The critical current density in full penetration can be estimated using the critical state model as: $j=20\Delta M/a(1-a/3b)$, where $\Delta M$ is the width of the magnetization hysteresis loop.

\begin{figure}[tb]   
\includegraphics*[bb=12 6 583 835, scale=0.34, angle=90]{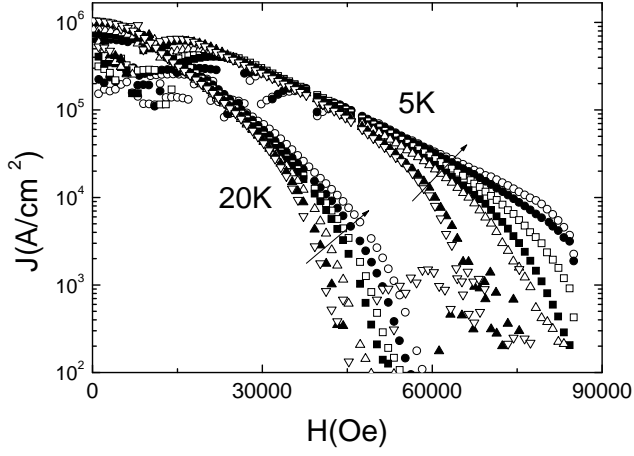}
\caption{The magnetic field dependent current density of the MgB$_2$ samples at 5 K and 20 K. The arrows indicate the direction of increasing sample size.} \label{jb}
\end{figure}

\begin{figure}[tb]   
\includegraphics*[bb=12 6 583 835, scale=0.34, angle=90]{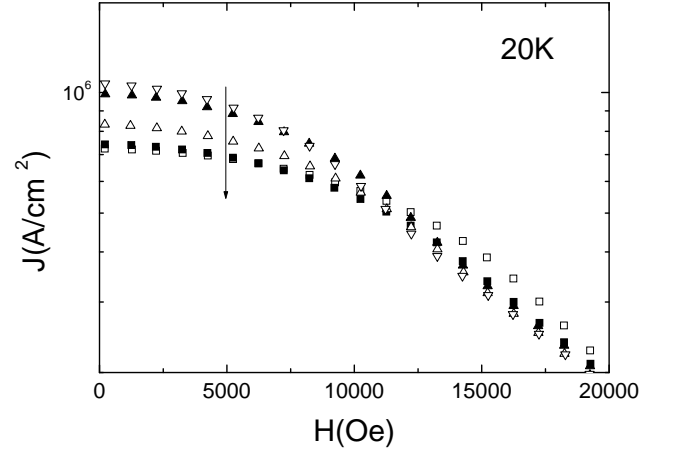}
\caption{The magnetic field dependent current density of the MgB$_2$ samples at low magnetic field for 5 K and 20 K. The arrows indicate the direction of increasing sample size.} \label{lowjb}
\end{figure}

Fig.\ \ref{jb} shows the magnetic field dependent critical current densities of all the samples at 5 K and 20 K. The arrows indicate the direction where the sample size increases. Flux jumping is observed for all samples at 5 K, but only for the two largest samples at 20 K. It can be clearly seen from Fig.\ \ref{jb} that at high fields (larger than 3 T) the current density increases systematically as the sample size increases. But the current density tends to be saturated when the sample size is very large. The low field part of Fig.\ \ref{jb} is enlarged in Fig.\ \ref{lowjb}, and again the arrow indicates the direction where the sample size increases. Contrary to what is observed at high fields, the current density decreases as the sample size increases.

In order to explain this observation, we start from the flux creep equation derived from Maxwell's equation $\nabla\times {\mathbf E}=-\partial{\mathbf B}/\partial t$ with ${\mathbf E}={\mathbf B}\times {\mathbf v}$ as discussed by Jirsa et al \cite{jirsa} and Schnack et al. \cite{schnack},
\begin{equation}
-\frac{dM}{dt}=\frac{\chi_0}{\mu_0}\frac{dB_e}{dt}-\frac{\Delta\nu_0B_e}{\mu_0}\exp\left[-\frac{U(j)}{kT}\right]
\label{creep}
\end{equation}
The vortex velocity $\nu$ is assumed to be thermally activated, i.e. $\nu=\nu_0\exp[-U(j)/kT]$, where the attempt velocity $\nu_0=x_0\omega_0$ with attempt frequency $\omega_0$ and attempt distance $x_0$ is the velocity of vortices when $U(j)=0$(i.e. $j=j_c$). $U(j)$ is the activation energy and $k$ the Boltzmann constant. The differential susceptibility $\chi_0$ and the geometric factor $\Delta$ in Eq.\ (\ref{creep}) depend on the size and shape of the sample. Here we consider a disk with $B_e$ parallel to its axis, then we have
\[
\chi_0=\frac{\pi^2R^3}{3\Im}
\]
\[
\Delta=\frac{2\pi^2R^2}{3\Im}
\]
where $R$ is the radius of the disk, $\mu_0R\Im$ the self-inductance of the disk. Eq.\ (\ref{creep}) can be solved for $U(j)$ as
\begin{equation}
U(j)=kT\ln\left[\frac{B_e\Delta\nu_0}{\mu_0\frac{dM}{dt}+\chi_0\frac{dB_e}{dt}}\right]
\label{uj}
\end{equation}

For a hysteresis loop measurement, we usually have $\mu_0\frac{dM}{dt}<<\chi_0\frac{dB_e}{dt}$, and Eq.\ (\ref{uj}) can then be reduced to
\begin{equation}
U(j)=kT\ln\left[\frac{B_e\Delta\nu_0}{\chi_0\frac{dB_e}{dt}}\right]=kT\ln\left[\frac{2B_e\nu_0}{R\dot{B}_e}\right]
\label{ujshort}
\end{equation}
here $\dot{B}_e=dB_e/dt$. Eq.\ (\ref{ujshort}) is simply related to the current-voltage $I-V$ curves (or $j$ versus $E$ curves where $j$ is the current density and $E$ the electric field) since for a cylinder of radius $R$, Faraday's law leads to
\begin{equation}
E=\frac{R}{2}\frac{dB_e}{dt}
\label{ej}
\end{equation} 
As can be seen from Eq.\ (\ref{ej}), a larger sample size $R$ will lead to a larger electric field in the sample, and therefore to a larger current density in the sample. This effect is similar to the effect of $\dot{B}_e$ on the hysteresis loop as has been used in dynamical relaxation measurements \cite{jirsa,schnack}.

According to Eq.\ (\ref{ujshort}), a different $U(j)$ will result in a different dependence of the current density on the sample size. The relaxation results of MgB$_2$ samples have led to a logarithmic dependence of the activation energy on the current density \cite{qin,han}
\begin{equation}
U(j)=U_0\log\frac{j_c}{j}
\label{log}
\end{equation}
where $U_0$ is the energy scale and $j_c$ the true critical current density at which $U(j_c)=0$.

Combining Eq.\ (\ref{ujshort}) with Eq.\ (\ref{log}), we obtain the sample size dependence of the current density as
\begin{equation}
j=j_cR^{1/n}\left(\frac{\dot{B}_e}{2B_e\nu_0}\right)^{1/n}
\label{ja}
\end{equation}
where $n=U_0(B,T)/kT$. The above analysis can also be applied to a rectangular rod with $R\approx ab/(a+b)$.

\begin{figure}[t]   
\includegraphics*[bb=12 6 583 835, scale=0.34, angle=90]{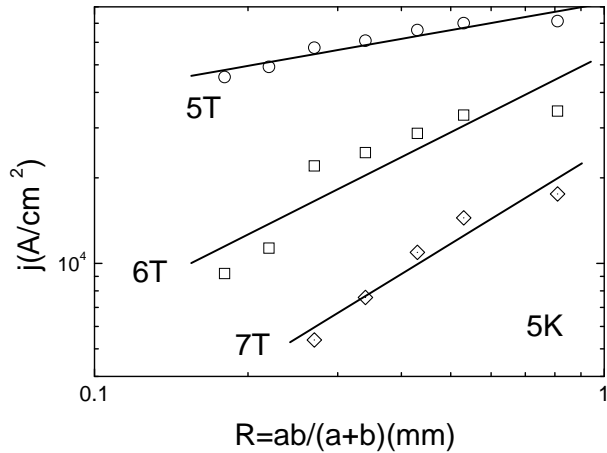}
\caption{The sample size dependence of the current density of the MgB$_2$ samples at 5 T, 6 T and 7 T for 5 K. Solid lines are the best fitting to the linear dependence.}
\label{jr5k}
\end{figure}

As can be seen from Eq.\ (\ref{ja}), the current density depends on the sample size as $j\propto R^{1/n}$, therefore the dependence is determined by the exponent $n$, which is a function of both temperature $T$ and magnetic field $B$. If $n$ is very large, there will be no sample size dependence as $R^{1/n}\to 1$, and this might be the reason why no sample size dependent current density has been reported in low temperature superconductors, characterizing the $E-j$ curve of the superconductors $E=E_c(j/j_c)^n$.  Typical $n$ factors in low temperature superconductors vary between 10 and 100 \cite{seeber}. As an example, magnets which work in the persistent mode without a drift require wires with a high $n$ factor, typically larger than 30 at the highest field \cite{seeber}. In high temperature superconductors, $n$ values as low as 5 in NdBa$_2$Cu$_3$O$_7$ \cite{inoue} and 4 YBa$_2$Cu$_3$O$_7$ \cite{inoue2} at high temperature and high magnetic field have been reported, indicating that a significant sample size dependent current density should be observed. However, weak-links in polycrystalline high temperature superconductors are very serious and prevail against the effects shown in Eq.\ (\ref{ja}), resulting in a lower critical current density in larger samples. 

Although the activation energy was reported to be very high in magnesium diboride at low temperature and low magnetic field, it drops sharply as the applied magnetic field and the temperature are increased \cite{qin}. The $n$ factors in MgB$_2$/Fe tapes and wires have been reported \cite{suo, beneduce} to be around 60 at 4 T, but drop to below 10 at high fields. From the $I-V$ curves of MgB$_2$ high density bulk samples reported by Pradhan et al. \cite{pradhan}, the $n$ factor is derived to be around 1.5 at 26.5 K and 5 T. A similar $n$ factor around $1$ has been obtained from $I-V$ curves by Kim et al. \cite{kim} at 30 K and 3 T. When the $n$ factor is in this range, the sample size dependence of the current density is expected to appear as seen in Fig. 1. The power law dependence $R^{1/n}$ saturates as $R$ is increased if $n$ is larger than $1$, which explains the reported saturation of the current density. Although no saturation is expected at very high temperatures and fields ($n$ may drop below 1), the total current is limited by the irreversibility line as $B\to B_{irr}$

\begin{figure}[t]   
\includegraphics*[bb=12 6 583 835, scale=0.34, angle=90]{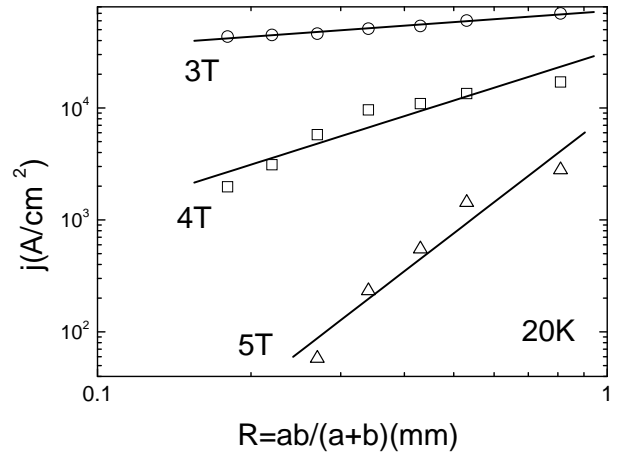}
\caption{The sample size dependence of the current density of the MgB$_2$ samples at 3 T, 4 T and 5 T for 20 K. Solid lines are the best fitting to the linear dependence.}
\label{jr20k}
\end{figure}

\begin{figure}[tb]   
\includegraphics*[bb=12 6 583 835, scale=0.34, angle=90]{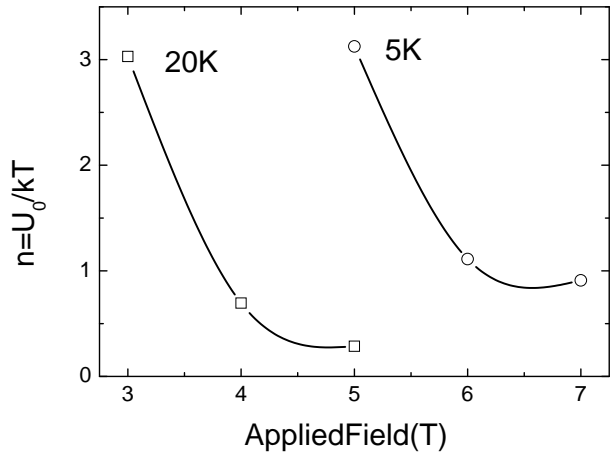}
\caption{The $n$ factor as a function of the applied magnetic field of the MgB$_2$ samples at 5 K and 20 K. The solid lines are only guides to the eyes.}
\label{nb}
\end{figure}

Eq.\ (\ref{ja}) provides us with a new method to determine the $n$ factor $n=U_0(B,T)/kT$ by plotting $\ln j$ versus $\ln R$. The inverse of the slope is just $n$. This method is applied to the samples used in this study, the results are shown in Fig.\ \ref{jr5k} and Fig.\ \ref{jr20k}, where $\ln j$ versus $\ln R$ curves at 5 T, 6 T, and 7T at 5 K, and 3 T, 4 T, 5 T at 20 K are plotted respectively. The solid lines in Fig.\ \ref{jr5k} and Fig.\ \ref{jr20k} are the best fittings to the linear dependence between $\ln j$ versus $\ln R$. The derived $n$ factor is shown in Fig.\ \ref{nb} as a function of the applied magnetic field at 5 K and 20 K. The solid lines are only guides to the eyes. This will be a very useful method when the critical current is too high to construct the whole $I-V$ curve to derive the $n$ factor.

On the other hand, as the $n$ factor is very high at low fields  \cite{suo, beneduce} (more than 100), a sample size dependence of the current density is not expected ($R^{1/n}\to 1$). The decreasing current density as the sample size increases might result from the self-field effect. Because larger samples carry larger currents (even if current density is almost the same), generating a larger self-field, this results in a smaller current density. Significant self-field has been observed in high temperature superconductors, especially in tapes with large critical currents \cite{schwartzkopf,spreafico}. And MgB$_2$ is expected to show similar behavior. Another possible reason is due to the surface pinning effect \cite{bean}. In the presence of both bulk and surface pinning, the magnetization is just the sum of the bulk and surface contributions \cite{burlachkov}. However, the surface component is only effective in the fields $H<H_p\approx \kappa H_{c1}/\ln\kappa$, with $\kappa$ the Ginsburg-Landau parameter and $H_{c1}$ the lower critical field \cite{burlachkov}. For the rectangular rods in this study, the ratio between the surface area parallel to the applied magnetic field and the sample volume is $2(ac+bc)/abc=2/R$, indicating a larger surface contribution as the sample size is decreased. This results in a larger current density in a smaller sample. The sample size dependence of the current density at low magnetic fields will be studied in more detail in our forthcoming work.

Another advantage of Eq.\ (\ref{ja}) is that we can use it to determine the current density dependent activation energy $U(j)$ in the sample. This is because different $U(j)$ relationships lead to different sample size dependences. For example, the linear current density dependent activation energy $U(j)=U_0(1-j/j_c)$ will give rise to a logarithmic dependence of the current density on the sample size,
\begin{equation}
j=j_c\left[1+\frac{1}{n}\ln a+\frac{1}{n}\ln\frac{\dot B_e}{2B_e\nu_0}\right]
\end{equation}
which is different from the power law dependence shown in Eq.\ (\ref{ja}). The experimental results shown in Fig.\ (\ref{jr5k}) and Fig.\ (\ref{jr20k}) indicate that in magnesium diboride superconductors, the activation energy depends logarithmically on the current density (Eq.\ (\ref{log})), rather than having the linear Anderson-Kim type dependence suggested by Jin et al. \cite{haojin}.

In summary, the dependence of the current density on the sample size in magnesium diboride superconductors has been observed and explained in this Letter based on the electric field generated in the superconductors. Starting from the flux creep equation, we have derived an analytical expression $j\propto R^{1/n}$ for the sample size dependent current density. We have shown that the sample size dependence of the current density can be used to derive the $n$ factor of MgB$_2$ samples and can also be used to determine the dependence between the activation energy and the current density. 

The authors gratefully acknowledge helpful discussions with D. C. Larbalestier and E. W. Collings. The authors are also very grateful to HyperTech Research Inc., Alphatech International Ltd., and the Australian Research Council for financial support.

\end{document}